\documentclass[11pt]{article}
\usepackage{amsmath,amsfonts,amsthm,amssymb,amscd}
\usepackage[all]{xy}
 
\begin{document}
\author {{\bf N. Bazunova$^*$, A. Borowiec$^{**}$ and R. Kerner$^{\#}$}}
\title{{\bf Universal differential calculus on ternary algebras}}
\maketitle
\date{ %}
%N. Bazunova, A. Borowiec and R. Kerner) \\
$^*$: Institute of Mathematics, University of Tartu, Tartu, Estonia \\
$^{**}$ Institute for Theoretical Physics, University of Wroc{\l}aw, Wroc{\l}aw, Poland \\
$^{\#}$ LPTL, Universit\'e Paris-VI, CNRS UMR 7600, 4 Place Jussieu, Paris, France}
%%%%%%%%%%%%%%%%%%%%% definitions %%%%%%%%%%%%%%%%%%%%
\def\me{m_{\alpha}\otimes e^{\alpha}}
\def\Todd{T^{\textrm {odd}}V}
\def\ootimes{\overline{\circledast}}
\def\Ua{\mathcal{U}_{\mathcal{A}}}
\def\Um{\mathcal{U}_{\mathcal{M}}}
\def\cd{\circledast}
\def\dt{\tilde{d}}
\def\Ac{\mathcal{A}}
\def\Mc{\mathcal{M}}
\def\OmegaU{\Omega _{u} ^1(\hat{\Ac})}
\newtheorem{example}{Example}
%%%%%%%%%%%%%%%%%%%%%%%%%%%%%%%%%%%%%%%%%%%%%%%%%%
\begin{abstract}
General concept of {\it ternary algebras} is introduced in this
   article, along with several examples of its realization. Universal
   envelope of such algebras is defined, as well as the concept of
   {\it tri-modules} over ternary algebras.
    The universal differential calculus on these structures is then
   defined and its basic properties investigated.\\
    {\it MSC2000:} 46L87, 58B34, 20N10.\\
    {\it Key words:} ternary algebras, tri-modules, universal ternary differential.
\end{abstract}
 \section{Introduction}

We start introducing some notation and conventions.
Throughout this article, we shall work in the category of vector spaces
over a field $\mathbb{K}$, which in our case, for simplicity,  is assumed to
be the field of real or complex numbers. This means that all objects considered
here are linear spaces, all mappings are ${\mathbb{K}}$--linear mappings,
the tensor product $\otimes$ is a shortcut for $\otimes_{\mathbb{K}}$.
Algebras will be generically denoted by ${\cal{A}}$, and the modules
will be denoted by ${\cal{M}}$.

In this letter we are interested in {\it ternary} algebras, i.e.
linear spaces over $\mathbb{K}$ endowed with a trilinear associative
composition law. More general structures of this type, called {\it n-ary algebras}
have been studied elsewhere (\cite{BDD}, \cite{Carls-2}, \cite{Connes},
\cite{DV-K2}, \cite{gelfand}, \cite{Karoubi}, \cite{kerner}, \cite{OPR}, \cite{vainerman}), and it has been shown that many familiar notions
from the theory of usual (i.e. "binary") algebras, such as
nilpotency, solvability, simplicity algebras etc., can be quite naturally
generalized to the $n$-linear case.

Our attention will be focused on particular properties of {\it ternary}
algebras, including the relations existing between general ternary
algebras or ternary algebras of particular types, and trivial
ternary algebras induced by the associative law in ordinary algebras, which
then play a role similar to the role played by associative algebras with
respect to the classical non-associative Lie algebras. Next, we shall
define the analog of {\it modules} over ordinary algebras, which  will be
called {\it tri-modules} in the present case.

Finally, we shall define the {\it derivations} of ternary algebras and
show in several examples how such differential ternary algebras can
be realized. General construction of the universal envelope for associative
ternary algebras and the universal differential calculus will also be
presented.

\section{Ternary algebras and tri-modules}
\subsection{Associative ternary algebras}

By {\it ternary} (associative) algebra $(\mathcal{A},[\ ])$ we mean a linear space
$\mathcal{A}$ (over a field $\mathbb{K}$) equipped with a linear map
$[\ ]:\mathcal{A}\otimes
\mathcal{A}\otimes \mathcal{A}\rightarrow \mathcal{A}$ called a (ternary)
multiplication (or product), which satisfies the following strong
associativity condition :
\begin{displaymath}
[[abc]de]=[a[bcd]e]=[ab[cde]]
\end{displaymath}
Weaker versions of ternary associativity , when only one of the
above identities is satisfied, can be called {\it left}
(respectively, {\it right} or {\it central}) associativity.

We look at associative ternary algebras as a natural
generalization of binary one: If $(\Ac, \cdot)$ is the usual (binary,
associative) algebra then an induced ternary multiplication can
be, of course, defined by $[abc]=(a\cdot b)\cdot c=a\cdot
(b\cdot c)$. In what follows, such ternary algebras will be called
trivial; from now on we shall study exclusively non-trivial ternary
algebras. It is known that unital ternary algebras are trivial.
Later on we shall  show that any finitely generated ternary algebra
is a ternary subalgebra of some trivial ternary algebra, which
is a ternary generalization of Ado's theorem for finite-dimensional
Lie algebras.

As we have already mentioned, many notions known in the binary case can be
directly generalized to the
ternary case. For example, the notion of ternary $\star$--algebra is defined
by $[abc]^{*}=[c^{*} b^{*} a^{*}]$, where  the star operation
$* : \Ac \rightarrow \Ac$  is, as it should be, (anti-) linear
anti-involution which means $(a^*)^*=a$ and $(ab)^*=b^*a^*$.
By the way, the very concept of involution can be generalized so that it
becomes adapted to ternary structures. A ternary involution should satisfy
$((a^*)^*)^*=a$, as an example, we can introduce the operation $*$ such that
$[abc]=[b^*c^*a^*]$.
In some applications, an  important role is played by ternary algebras with
a different associativity law:
\begin{eqnarray}
\label{assB}
[[abc]de]=[a[dcb]e]=[ab[cde]].
\end{eqnarray}
Such associativity is sometimes called  ``type $B$-associativity''
or ``2nd kind'' \cite{Carls-1}. In the case of ternary
$\star$--algebras both types of associativity are related to each
other. Assuming that $(\Ac, [\ ],*) $ is a ternary $\star$--algebra,
one can introduce another ternary multiplication ${[\ ]}_{*}$ such
that
$$
{[abc]}_{*}\stackrel{def}{=} [a b^{*} c], \quad \forall \ a,\ b,\ c \in \Ac .
$$
The algebra $(\Ac, {[\ ]}_{*}, \,* )$ becomes an associative
ternary $\star$--algebra of $B$-type. The converse statement is also true:
any ternary $\star$--algebra of $B$-type gives rise to a standard
ternary $\star$--algebra. Observe that in the case of
algebras over the field of complex numbers one has to assume
anti-linearity of ternary multiplication in the middle factor
instead of linearity.\\

{\bf Example} Any Hilbert or symmetric scalar product (i.e. metric) vector space
$\mathcal{H}$ bears a canonical
structure of ternary algebra of $B$-type with ternary
multiplication defined as follows:
$$\{a\, b\, c\}= <a,\,b>\,c$$
induced by  scalar multiplication $<,\,>$ in $\mathcal{H}$. It is not a
ternary $\star$--algebra.
%And the algebra $(\Ac , [\ ])$ with operation $*$ becomes associative of type $B$.
%--------------\\

In the finite-dimensional case, when we replace the Hilbert
space with a metric vector space, in a given basis $\{e_k\}, \,
k;m = 1,2 \dots N$ we can define a non-degenerate metric $g_{ik} =
<e_i , e_k>$. Then the Clifford algebra generated by the elements
$C_i$ satisfying
$$C_i C_j + C_j C_i = 2 g_{ij} {\bf 1} $$
provides an appropriate associative algebra which can serve as a
representation of the ternary product $\{e_i e_j e_k\} = <e_i
, e_j>e_k$ as follows:
$$ \{C_i\,  C_j\,  C_k \} = \frac{1}{2} \, (C_i C_j + C_j C_i) \, C_k$$
\indent Obviously, there are two other possible choices of ternary
product in a metric (or Hilbert) space, corresponding to cyclic
permutations of three factors:
$$\{ a\, b\, c \}' = <b,\,c>\, a , \, \ \ \, \{a\,b\,c\}" = <c,a> b$$
In a finite-dimensional case, one may define the most general
ternary product of this type as a linear combination of these
three, i.e.
\begin{equation}
\{ e_i\,  e_j\,  e_k \} = \displaystyle{\sum_{l,m,n} \, M^{lmn}_{ijk}
\, <e_l ,\, e_m> \, e_n } = \rho^{n}_{\, \, ijk} \, e_n
\end{equation}
with tensors $M^{lmn}_{ijk}$ symmetric in first two upper indices
$l,m$. The four-index tensor $\rho^n_{\, \, ijk}$ plays the role
of structure constants of our ternary algebra.
It is easy to prove that it is impossible to impose strong
associativity on such a product, because the set of equations it
would imply on the coefficients of the tensor $M^{lmn}_{ijk}$ is
strongly over-determined (see (\cite{vainerman}) for example).

As in the usual algebraic case, we can impose particular symmetries
on the ternary product, defining a new product displaying
a representation property with respect to the permutations of
its three lower indices, e.g. by requiring the total symmetry:
$$\{e_i\, e_j\, e_k \}_{\rm sym} = \{ e_i\, e _j\, e_k \} + \{ e_j\, e_k\, e_i \}
+  \{e_k\, e_i\, e_j \}  =$$
\begin{equation}
<e_i, e_j> \, e_k + <e_j, e_k> \, e_i + <e_k, e_i> \, e_j
\end{equation}
Other choices are possible; for example, a $Z_3$-generalization of
the commutator in associative binary algebras, which generates
non-associative Lie algebras, can be introduced as follows:
\begin{equation}
\{e_i\, e_j\, e_k \}_q = \{ e_i\, e_j\, e_k \} + q \, \{ e_j\, e_k\, e_i \}
+  q^2 \{e_k\, e_i\, e_j \}
\end{equation}
with $q$ one of the primitive third roots of unity, $q = e^{\frac{2 i \pi}{3}}$,
satisfying $q^3 = 1$ and $q + q^2 + q^3 = 0$.

This algebra is particularly simple in dimension two, when there are only
two basis vectors. Then the "ternary structure constants" $\rho^i_{\, jkm}$
are as follows:
$$ \rho^i_{\, 111} = \rho^i_{\, 222} = 0;$$
$$ \rho^1_{\, 221}=q \, \rho^1_{\, 212} = q^2 \, \rho^1_{\, 122} =1;
\, \ \ \,  \rho^2_{\, 112}=q \, \rho^2_{\, 121} = q^2 \, \rho^2_{\, 211}
 = 1  $$
\begin{equation}
 \rho^2_{\, 221}=q \, \rho^2_{\, 212} = q^2 \, \rho^2_{\, 122} =0 ;\, \ \ \,
\rho^1_{\, 112}=q \, \rho^1_{\, 121} = q^2 \, \rho^1_{\, 211}  = 0
\end{equation}
Besides their particular symmetry, the coefficients $\rho^i_{\, jkm}$
possess  another interesting property akin to the representation property
of antisymmetric structure constants of usual Lie algebras. Let us introduce
the following ternary composition law for these coefficients, which can be
also named "cubic matrices", with regard to their lower indices:
\begin{equation}
(\rho^i * \rho^j * \rho^k)_{prs} = \displaystyle{\sum_{nmt}} \,
\rho^i_{\, npm} \, \rho^j_{\, mrt} \, \rho^k_{\, tsn}
\end{equation}
Then, introducing the same $Z_3$-skew-symmetric product as
\begin{equation}
\{ \rho^i \, \rho^j \, \rho^k \}_q = (\rho^i * \rho^j * \rho^k) +
q \, (\rho^j * \rho^k * \rho^i) + q^2 \, (\rho^k * \rho^i * \rho^j) ,
\end{equation}
we can easily check that
\begin{equation}
\{ \rho^i \, \rho^j \, \rho^k \}_q =
\displaystyle{\sum_{m}}  \rho_m^{\, ijk} \, \rho^m
\end{equation}
which provides us with a faithful representation of our ternary algebra.
\newline
\indent
As in the usual case, we are interested to know whether such an algebra
can be also represented by certain combinations of ternary products in
an ordinary (i.e. binary) associative algebra, playing the role of an
enveloping algebra. It is easy to see that the answer is positive. In the
above example, the two generators of non-associative ternary  algebra with
$Z_3$-skew-symmetric product can be represented by any two Pauli matrices
multiplied by the factor $i/2$. One can check that the matrices
$\tau_i = \frac{i}{2} \, \sigma_i \, , \, \ \ i = 1,2 \,  $ satisfy
\begin{equation}
\sigma_i \sigma_j \sigma_k + q \, \sigma_j \sigma_k \sigma_i
+ q^2 \, \sigma_k \sigma_i \sigma_j = \displaystyle{\sum_m}  \,
\rho^m_{\, ijk} \, \sigma_m
\end{equation}
%%%%%%%%%%%%%%%%%%%%%%%%%%%%%%%%%%%%%%%%%%%%%%%

\subsection{Universal envelope of ternary algebra}

In order to construct an universal envelope of ternary algebra
one should consider the structure of ternary algebras in more detail.
In the classical (binary) case, algebras defined by generators and
relations between them play important role in concrete applications.
It suffices to mention that the well-known Grassmann and Clifford
algebras can be defined in this way.
Let us briefly summarize this approach (see e.g. \cite{Bour}).
First, we recall that the tensor algebra
$$TV = \oplus _{k=0}^{\infty} V^{\otimes k}= \mathbb K\oplus
V\oplus V^{\otimes 2}\oplus V^{\otimes 3} \oplus \dots \ $$
of a given vector space $V$ is a  free algebra with $n$-generators,
where $n={\rm dim}\,V$. Thus for any subset $S\subset TV$ one can construct
a two-sided ideal $J_S$ generated by $S$ and the quotient algebra
$$\Ac_S = TV / J_S\ .$$
Here $V$ is called a space of generators, $S$ is a set of
generating relations. Conversely, by the well-known  theorem
(see e.g. \cite{Bour}) any (unital)
algebra with $n$- generators can be obtained in this way.
Notice that a non-unital free algebra can be defined as
$$T^\prime V = \oplus _{k=1}^{\infty} V^{\otimes k}=
V\oplus V^{\otimes 2}\oplus V^{\otimes 3} \oplus \dots \ .$$

Much in the same way, for any vector space $V$ one can construct a free
ternary algebra generated by $V$. To this aim we define
\begin{eqnarray}
\Todd =\oplus _{k=0}^{\infty} V^{\otimes (2k+1)}=
V\oplus V^{\otimes 3}\oplus V^{\otimes 5} \oplus \dots \ .
\end{eqnarray}
as a ternary algebra with a ternary multiplication:
\begin{eqnarray}
{[uvw]}_{\otimes}=u\otimes v\otimes w, \quad \forall\ u,\ v,\ w \in \Todd\ .
\end{eqnarray}
Observe that $\Todd$ is not a trivial ternary algebra, however it is a ternary
subalgebra in the trivial ternary algebra $TV$ (as well as in $T^\prime V$).

$\Todd$ plays the role of free ternary algebra in the following way.
Let $(\mathcal{A}, [\ ])$
be a ternary algebra, $V$ any vector space. Then for any linear map
$\varphi: V \rightarrow \mathcal{A}$, there
exists its unique lift
$\tilde{\varphi}: \Todd \rightarrow  \mathcal{A}$,
which is a homomorphism of ternary algebras, such that
$\varphi=\tilde{\varphi} \circ \mu$, which means that
the following diagram
$$
\xymatrix{
                 &  \Todd \ar[d]^-{\tilde{\varphi}}\\
\ar@{^{(}->}[ur]^-{\mu} V
\ar[r]^-{\varphi}
& \mathcal{A}}
$$
is commutative. Here $\mu$ denotes the canonical embedding
${\mu}: V\hookrightarrow\Todd$.

In particular, if $\varphi$ is an embedding and $\tilde{\varphi}$ is an
epimorphism, then
$$\mathcal{A}\cong \Todd / Ker (\tilde{\varphi})$$
i.e. the algebra $\mathcal{A}$ becomes isomorphic to the quotient algebra
$\Todd / Ker (\tilde{\varphi})$.
Quite obviously, ${\rm Ker}\tilde {\varphi}$ is a ternary ideal in $\Todd $.
The simplest example is provided tautologically by the fact that
%\mbox{($<-$ control spelling)}
$$\mathcal{A}\cong T^{\textrm{odd}} \Ac /gen<a\otimes b\otimes c - [abc]>,$$
where $gen<a\otimes b\otimes c - [abc]>$ denotes a ternary ideal generated by
elements $\{a\otimes b\otimes c - [abc]:\ a,\ b,\ c \in \Ac \}$.

Any $n$-nary algebra can be embedded into a
binary one \cite{Carls-2}. Here we are particulary interested in the
ternary case \cite{Carls-1}. For given ternary algebra  $\mathcal{A}$
one can defined  a $\mathbb{Z}_2$--graded vector space
\begin{eqnarray}
\Ua=
\mathcal{A}_1\oplus \mathcal{A}_0, \nonumber
\end{eqnarray}
where $\mathcal{A}_1=\mathcal{A}$ is an odd part. The even subspace
$\mathcal{A}_0$ of $\Ua$ is assumed to be the quotient vector space $$\mathcal{A}_0=
(\mathcal{A}\otimes \mathcal{A})/{\rm span}<[xyz]\otimes w - x\otimes [yzw]>$$
where ${\rm span}<[xyz]\otimes w - x\otimes [yzw]>$ denotes a vector subspace
of $\Ac\otimes\Ac$ span by elements
$\{[xyz]\otimes w - x\otimes [yzw]: x,y,z,w\in\Ac\}$. Let $a\circledast b$ denote
the equivalence class of the element $a\otimes b\in \mathcal{A}\otimes \mathcal{A}$.
Now we are in a position to define the multiplication  $\ootimes$
between elements from $\Ua$ by the following:
\begin{eqnarray}
\begin{array}{c}
a\ootimes b \stackrel{def}{=} a\circledast b;\nonumber \\
(a\circledast b) \ootimes c = a\ootimes (b \circledast c)
\stackrel{def}{=} [abc]; \nonumber \\
(a\circledast b) \ootimes (c\circledast d) \stackrel{def}{=} [abc]\circledast d =
a\circledast [bcd]=a\ootimes(( b\cd c)\ootimes d)=
 (a\ootimes( b\cd c))\ootimes d\ .
\nonumber
\end{array}
\end{eqnarray}
In this way, we have obtained a $\mathbb{Z}_2$-graded  algebra, since
$$\mathcal{A}_i \circledast \mathcal{A}_j \subset \mathcal{A}_{i+j(mod2)}.$$
It is easy to see that this binary, nonunital algebra %$\Ua$
is associative. The initial ternary algebra $\Ac\equiv\Ac_1$ becomes
a ternary subalgebra in the trivial ternary algebra $\Ua$.
Of course, $\Ac_0$ is a (binary) algebra which is also a subalgebra of $\Ua$,
and $\Ac$ becomes a $\Ac_0$--bimodule.
Further on, to simplify the notation, we shall use the same symbol
in order to denote the equivalence class
$a\circledast b \in \mathcal{A}_0$ corresponding to the elements
$a,\ b \in \mathcal{A}$, and for the multiplication $\ootimes$
in $\Ua$.

In other words, any ternary algebra can be extended to a binary one, i.e.
a ternary algebra is a ternary subalgebra in some associative binary algebra.
Furthermore, if $\phi : \Ac \rightarrow \mathcal{B}$ is a ternary homomorphism
of $\Ac$ into an associative algebra $\mathcal{B}$, i.e.
$\Ac$ is a ternary subalgebra in a binary algebra $\mathcal{B}$,
%and $\phi : \Ac \rightarrow \mathcal{B}$ denotes the corresponding embedding,
then there exists one and only one (binary) algebra homomorphism
$\tilde{\phi} : \Ua \rightarrow \mathcal{B}$ such that $\phi=\tilde{\phi}\circ \iota$,
where $\iota$ is the canonical embedding of $\Ac$ into $\Ua$, i.e. the
following  diagram:
$$
\xymatrix{
                 &  \Ua \ar[d]^-{\tilde{\phi}}\\
\ar@{^{(}->}[ur]^-{\iota} \Ac
\ar[r]^-{\phi}
& \mathcal{B}}
$$
is commutative, and this universal property characterizes  $(\iota \,,\Ua)$
up to an isomorphism. For example, $\mathcal{U}_{\Todd} =T^\prime V$, i.e.
\begin{eqnarray}
T^\prime V = V\oplus V^{\otimes 2} \oplus V^{\otimes 3}\oplus V^{\otimes 4}
\oplus \dots \ =\Todd \oplus T^{\textrm {even}}V,
\nonumber
\end{eqnarray}
is the enveloping algebra, in which the ternary algebra $(\Todd, {[\ ]}_{\otimes})$
can be embedded.

\subsection{Tri-modules over ternary algebras.}

The concept of tri-module is a particular case of the concept of module
over an algebra over an operad defined in \cite{GK}.
In the more general context of $n$-ary algebras it was then considered
in \cite{Gned}. Here, a structure of tri-module over a ternary algebra
$\mathcal{A}$ is simply defined on a vector space $\Mc$ by the following
three linear mappings called
{\center{
 left\ $\qquad\ [\ ]_{L}:\mathcal{A}\otimes \mathcal{A}\otimes
\mathcal{M}\rightarrow \mathcal{M}$,\\
 right\ $\qquad\ {[\ ]}_{R}:\mathcal{M}\otimes \mathcal{A}\otimes
\mathcal{A}\rightarrow \mathcal{M}$,\\
and central\ $\qquad\ {[\ ]}_{C}:\mathcal{A}\otimes \mathcal{M}\otimes
\mathcal{A}\rightarrow \mathcal{M}$\\ }}
\noindent
multiplication, respectively (see also \cite{Carls-1,Michor}). They are assumed to
satisfy the following compatibility conditions
\begin{eqnarray}
\label{lmod}
[ab[cdm]_{L}]_{L}=[[abc]dm]_{L}=[a[bcd]m]_{L}, \\
\label{rmod}
[[mab]_{R}cd]_{R}=[ma[bcd]]_{R}=[m[abc]d]_{R},\\
\label{cmod} [a[b[cmx]_{C}y]_{C}z]_{C}=[[abc]m[xyz]]_{C},\\
\label{lcmod}
[a[bcm]_{L}d]_{C}=[ab[cmd]_{C}]_{L}=[[abc]md]_{C},\\
\label{rcmod}
[a[mbc]_{R}d]_{C}=[[amb]_{C}cd]_{R}=[am[bcd]]_{C},\\
\label{lrmod}
[[abm]_{L}cd]_{R}=[ab[mcd]_{R}]_{L}=[a[bmc]_{C}d]_{C},\\
\forall\ a,\ b,\ c,\ d,\ x,\ y,\ z \in\mathcal{A},\ m
\in\mathcal{M}\nonumber
\end{eqnarray}

In the case of tri-module $\mathcal{M}$ over an algebra $\mathcal{A}$ of
type $B$ the conditions (\ref{lmod}),\ (\ref{rmod}),\
(\ref{lrmod}) remain  unchanged while (\ref{cmod}),\ (\ref{lcmod}),\
(\ref{rcmod}) have to be replaced correspondingly by
\begin{eqnarray}
\label{newcmod}
[a[b[cmx]_{C}y]_{C}z]_{C}=[[ayc]m[xbz]]_{C},\\
\label{newlcmod}
[a[cbm]_{L}d]_{C}=[[amb]_{C}cd]_{R} ,\\
\label{newrcmod}
[a[mbc]_{R}d]_{C}=[ab[cmd]_{C}]_{L},\\
\forall\ a,\ b,\ c,\ d,\ x,\ y,\ z \in\mathcal{A},\ m
\in\mathcal{M}.\nonumber
\end{eqnarray}

Much in the same way as binary algebra is a trivial ternary algebra, the notion of
tri-module generalizes the notion of bimodule. More exactly, one has

REMARK. Let $\Ac$ be a (binary) algebra and $\Mc$ a bimodule over it.
Thus defining $[abm]_L=a\cdot(b\cdot m)=(a\cdot b)\cdot m$, $[amb]_C=a\cdot(m\cdot b)= (a\cdot m)\cdot b$ and $[mab]_R=m\cdot(a\cdot b)= (m\cdot a)\cdot b$
we see that $\Mc$ becomes a tri-module over the same algebra considered as a trivial ternary algebra.
%%%%%%%%%%%%%%%%%%%%%%%%%%%%%%%%%%%%%%%%%%%%%%%%%%%%%

Analogously, we can obtain an enveloping module $\Um$ over the enveloping algebra
$\Ua$ of ternary module $\mathcal{M}$ over ternary algebra $\mathcal{A}$.
Let us denote by $\Um$ a $\mathbb{Z}_2$-graded vector space
\begin{eqnarray}
\Um=\mathcal{M}_1 \oplus \mathcal{M}_0,
\end{eqnarray}
where the odd part $\mathcal{M}_1 \equiv \mathcal{M}$. The even part is
defined as a quotient vector space
$$\mathcal{M}_0 =(\mathcal{A}\otimes \mathcal{M}\oplus \mathcal{M}\otimes
\mathcal{A})/lin<S>,$$
where $S$ is a set of  elements in $\mathcal{A}\otimes \mathcal{M}\oplus \mathcal{M}\otimes \mathcal{A}$
($a,\ b,\ c,\in \Ac, \ m\ \in \mathcal{M}.$)
\begin{eqnarray}
\begin{array}{c}
[abc]\otimes m - a\otimes {[bcm]_L},\nonumber\\
{[abm]}_L \otimes c - a\otimes {[bmc]}_C,\nonumber\\
{[amb]}_C \otimes c - a\otimes {[mbc]}_R,\nonumber\\
{[mab]}_R\otimes c - m\otimes [abc],\nonumber
\end{array}
\end{eqnarray}
which generate the subspace $lin<S>$.

As previously, denote  by $a\cd m$ or $m\cd a$ the corresponding equivalence
classes, elements of $\mathcal{M} _0$.
Define left and right multiplication $\ootimes$ between elements
from $\Ua$ and those from $\Um$ in the following way:
\begin{eqnarray}
\begin{array}{c}
a\ootimes m \stackrel{def}{=}a\cd m ; \nonumber\\
m\ootimes a\stackrel{def}{=}m\cd a;\nonumber\\
(a\cd b)\ootimes m=a\ootimes (b\cd m)\stackrel{def}{=}[abm]_L ;\nonumber\\
m \ootimes (c\cd d)=(m \cd c) \ootimes d \stackrel{def}{=}[mcd]_R ;\nonumber\\
a\ootimes (m\cd b) \stackrel{def}{=}[amb]_{C} ; \qquad
(a \cd m) \ootimes b \stackrel{def}{=}[amb]_{C} ;\nonumber \\
\forall \ a,\ b,\ c,\ d,\ e,\ f\ \in \Ac, \ m\ \in \mathcal{M}.
\end{array}
\end{eqnarray}
One can check the following properties of  the action of algebra
$\mathcal{A}$ on module $\mathcal{M}$
%\begin{eqnarray}%\begin{array}{c}
$${[abc]} \cd m =a\cd {[bcm]}_L ; \nonumber $$
$$m\cd [bcd]  = [mbc]_R \cd d;\nonumber $$
$${[abc]} \cd (m \cd d) =[ab[cmd]_C]_L ;\nonumber$$
$$(a\cd m)\cd [bcd]  = [[amb]_C cd]_R ;\nonumber$$
$$(a\cd [bcd]) \cd m=({[abc]} \cd d) \cd m =
[abc] \cd (d \cd m)=[ab[cdm]_L]_L ;\nonumber$$
$$m\cd({[cde]}\cd f)=m\cd(c \cd [def])=
(m\cd c) \cd [def]=[[mcd]_R ef]_R ; \nonumber$$
$$(a\cd [bcd]) \cd (m \cd e)=([abc] \cd d) \cd (m \cd e)=[abc] \cd [dme]_C ;\nonumber$$
$$(a \cd m) \cd ({[bcd]}\cd e)= (a \cd m) \cd (b \cd [cde])
=[[amb]_C \cd [cde] ; \nonumber$$
%\end{array}
%\end{eqnarray}

Thus $\Um$ becomes a $\mathbb{Z} _2$-graded bimodule over $\Ua$ sinces
$$\Ac _i\ootimes\mathcal{M} _j \subseteq \mathcal{M} _{i+j(mod 2)}, \quad
\mathcal{M} _j\ootimes\Ac_i\subseteq \mathcal{M} _{i+j(mod 2)},\
\quad i,\ j \in \{ 0, 1\}.$$
In particular, $\Mc\equiv\Mc_1$ and $\Mc_0$ are $\Ac_0$-bimodules.

Further on, we shall use the same symbol $\cd$ to denote the equivalence
class in $\Um$, its bimodule structure and for the multiplication in $\Ua$.

Let us  stress again that any bimodule over a (binary) algebra becomes
automatically a trimodule over the same algebra considered as a trivial
ternary algebra.

What we have shown above is that any trimodule is a sub-trimodule of some universal
bimodule $\Um$ over $\Ua$. Conversely, if $\mathcal{N}$ is a
$\mathbb{Z} _2$--graded bimodule over
$\Ua$, then its odd part $\mathcal{N} _1$ is a trimodule over $\Ac$.
%%%%%%%%%%%%%%%%%%%%%%%%%%%%%%%%%%%%%%%%%%%%%%%%%%%%%%%%%%%
%%%%%%%%%%%%%%%%%%%%% SECTION 2 %%%%%%%%%%%%%%%%%%%%%%%%%%%
\section{ Universal differentiation of ternary algebra}
%\subsection{ First order differential calculus }

A first order differential calculus (differential calculus in short) of ternary algebra $\mathcal{A}$  is a linear map from ternary algebra a into tri-module
over it, i.e. $d: \mathcal{A}\rightarrow \mathcal{M}$, such that a ternary
analog of the Leibniz rule takes place:
\begin{eqnarray}
\label{terLeib}
d([f\,g\,h])=[df\,g\,h]_{R}+[f\,dg\,h]_{C}+[f\,g\,dh]_{L},\quad \forall
f,g,h \in \mathcal{A}.
\end{eqnarray}

In particular, if $\mathcal{M}=\Ac$, then we shall call so defined differential
{\it ternary derivation} of $\Ac$. An interesting example is provided by
\begin{example}
Ternary derivative  in Hilbert (or metric) vector space.\\
As we already noticed,
any Hilbert space $(\mathcal{H}, <,>)$ inherits a canonical
ternary 2nd type associative structure given by
$\{a\,b\,c\}=<a,b>c$. \\
For a linear operator being a ternary derivation
$D:\mathcal{H}\rightarrow\mathcal{H}$ one calculates:
% from the Leibniz rule for the derivative\\
\begin{eqnarray}
D\{a\,b\,c\}= \{Da\,b\,c \}+\{a\,Db\,c \}+ \{a\,b\,Dc \},
\nonumber
\end{eqnarray}
Now, taking into account that $D\{a\,b\,c\} = <a,b>Dc=\{a\,b\,Dc\}$ it implies
\begin{eqnarray}
<Da,b>=-<a,Db>\ \Rightarrow\ D^{+}=-D,\ \  i.e.\ \ (iD)^{+}=iD
\nonumber
\end{eqnarray}
i.e, that ternary derivations are in one-to-one correspondence
with hermitian operators in $\mathcal{H}$. This makes possible a link
with Quantum Mechanics, especially the version introduced by Nambu
(\cite{Nambu}).

\end{example}
Let us refer again to the classical (binary) case. First order differential calculus
from an algebra into bimodule can be automatically interpreted as
a ternary differential calculus from trivial ternary
algebra into a trivial tri-module over it. It can be easily seen from
$$d(fgh)=d((fg)h)=d(fg)\,h+fg\,d(h)=df\,gh
+f\,dg\,h+fg\,dh\ .$$
The converse statement is, in general, not true. A ternary Leibniz rule for
differential calculus from an algebra into bimodule does not necessarily imply,
in case of non--unital algebras, the existence of a standard
(binary) Leibniz rule. In particular, the set of ternary derivations
of non-unital algebra should be an extension of the set of standard
(binary) derivations.\\

Let $(\Ac, \Mc, d)$ be a our ternary differential calculus from
ternary algebra into tri-module. By the Leibniz rule
$$\tilde d (a\cd b)= (\tilde d a)\cd b + a\cd(\tilde d b)$$
it can be uniquely extended to a
$0$-degree differential $\tilde{d} : \Ua \rightarrow \Um$, in a way which
ensures commutativity of  the following diagram:
$$
\xymatrix{
\Ac\ar[d]^{\iota} \ar[r]^d
         &\Mc\ar[d]^{\iota^\prime}\\
\Ua\ar[r]^{\tilde d} & \Um }$$
Conversely, any $0$-degree  first order differential calculus from
$\Ua$ into $\Um$, such that $\tilde d\mid_{\Ac}\,\subset \Mc$ gives rise to ternary $\Mc$-valued  differential calculus on $\Ac$.\\

The universal  first order differential calculus on non-unital
algebras is well describe in \cite{Connes,Cu-Qu,Karoubi}.
Let us recall this construction shortly.
Determine a vector space
$\OmegaU =\hat{\mathcal{A}}\oplus \hat{\mathcal{A}}
\otimes \hat{\mathcal{A}}$, where $\hat{\mathcal{A}}$ is a non-unital
(binary) algebra.  Any element from $\OmegaU$
can be written in the form:
$(a,b\otimes c),\ {\rm where} \ a,\ b,\ c\ \in  \hat{\mathcal{A}}$.
Define left and right multiplications by elements from
$\hat{\Ac}$:
\begin{eqnarray}
x(a,b\otimes c)&=&(0, x\otimes a + xb\otimes c), \nonumber \\
(a, b\otimes c)y&=&(ay, -a\otimes y +b\otimes cy - bc\otimes y). \nonumber
\end{eqnarray}
In this way, $\OmegaU $ becomes a $\hat{\mathcal{A}}$-bimodule
since
$$ (x (a,b\otimes c)) y = x ((a,b\otimes c) y).$$

Let $D:\hat{\mathcal{A}}
\rightarrow \hat{\mathcal{A}}\oplus \hat{\mathcal{A}}\otimes \hat{\mathcal{A}},$
$Da= (a, 0),\quad \forall \ a \in \hat{\mathcal{A}}\ $ be a canonical embedding.
Because it satisfies the Leibniz rule:
$$
D(a)b+aD(b)=(a,0)b+a(b,0)=(ab,a\otimes b)+(0,-a\otimes b)=(ab,0)=D(ab),
$$
$D$ is a differential. We shall call it the universal differential for
a given algebra $\hat{\Ac}$.

For unital algebras, there exists an alternative construction of
$\OmegaU$ as a kernel of multiplication map \cite{Cart} (see also \cite{Bour}).
Since our ternary
algebras have no unit element, we  can not use such construction here.

Our aim is to provide an analogous construction in the case
of ternary algebra.

From a $\mathbb{Z}_2$-graded $\Ua$-bimodule $\Omega^1_u(\Ua)=
\Ua\oplus \Ua\otimes \Ua$, let us extract its odd subspace $\Ac\oplus
{\Ac}_0\otimes \Ac \oplus \Ac \otimes \Ac_0$ with elements
$$(a, \beta \otimes b, c\otimes \gamma),\ \forall\ a,\ b,\ c,\ \in \Ac,\
 \beta,\ \gamma \in \Ac _0 .$$
We shall denote it as $\Omega^1_T(\Ac)=\Ac\oplus
{\Ac}_0\otimes \Ac \oplus \Ac \otimes \Ac_0$. As we already know from
our previous considerations, $\Omega^1_T(\Ac)$ is a tri-module over $\Ac$.
Thus we have defined the left, central and right ternary multiplications
%by the following formulas respectively
\begin{eqnarray}
\begin{array}{l}
%\begin{eqnarray}
\label{trimult}
[xy(a,\beta \otimes b, c \otimes \gamma ) ]_L=
(0, (x\cd y) \otimes a + (x\cd [y \beta ]) \otimes b, [xyc] \otimes \gamma ); \\
\\
{[ x ( a, \beta \otimes b, c \otimes \gamma ) y  ]}_C =
(0, -(x\cd a)\otimes y - ([x\beta]\cd \gamma) \otimes y + (x\cd c) \otimes [\gamma y] -\\
\\
(x\cd [c\gamma ])\otimes y,
x\otimes (a\cd y)+ [x\beta] \otimes (b\cd y)) ;\\
\\
{[(a,\beta \otimes b, c \otimes \gamma )xy ]}_R=\\
\\
([axy], \beta \otimes [bxy], -a\otimes (x\cd y)-[\beta b]\otimes (x\cd y) +
c\otimes ([\gamma x]\cd y) - [c \gamma ] \otimes (x\cd y)).
\end{array}
\end{eqnarray}

%As it follows from a direct calculation, the properties (\ref{lmod} -- %\ref{lrmod}) take a place, and the module $\Omega ^1(\Ac)$ became a
%trimodule under multiplication (\ref{trimult}).

The canonical embedding $D:\Ac\rightarrow \Omega_T ^1{\Ac}$:
%by the formula:
\begin{eqnarray}\label{t-diff}
D(a)= (a,0,0), \quad \forall \ a \in \Ac.
\end{eqnarray}
defines a ternary differential (\ref{terLeib}). In fact, one has
\begin{eqnarray}
\begin{array}{l}
%[D(a) bc]_L +[a D(b) c]_C+[ab D(c)]_R=
[(a,0,0)bc]_L + [a(b,0,0)c]_C + [ab(c,0,0)]_R =\\ ([abc],0,a\otimes (b\cd c))+
(0,-(a\cd b) \otimes c,-a\otimes (b\cd c))+(0,(a\cd b)\otimes c,0)=\\ ([abc],0,0).
%D([abc]).
\end{array}\nonumber
\end{eqnarray}

This  ternary differential calculus is universal because for any trimodule $E$ and
any ternary $E$-valued  differential calculus   $d:\Ac\rightarrow E$, there
exists one and only one covering trimodule homomorphism ${\tilde{\varphi}}_d$
such that $d={\tilde{\varphi}}_d \circ D$, i.e. the following diagram
$$
\xymatrix{
                 &  {\Omega^1_T(\Ac)}  \ar[d]^-{{\tilde{\varphi}}_d}\\
\ar@{^{(}->}[ur]^-{D} \Ac
\ar[r]^-{d}
& E}
$$
is commutative.
Moreover, if the trimodule $E$ is spanned by the elements $d\Ac ,\ [\Ac \ d\Ac\ \Ac ]_C$ and $[d\Ac\ \Ac\ \Ac]_R$, then ${\tilde{\varphi}}_d$ is an epimorphism and
$E=\Omega_T^1(\Ac)/Ker(\tilde{\varphi} _d)$.

In this way, the problem of classification of all first order differential
calculi over $\Ac$ can be translated into the problem of classification of all
sub-trimodules in $\Omega^1_T(\Ac)$. Remember  that  $\Omega^1_T(\Ac)$ is an
odd part of $\Omega_{u} ^1 (\Ua)$ and our ternary differential (\ref{t-diff})
is, in fact, a restriction of the universal differential.

As it is well known \cite{Connes,Cu-Qu,Karoubi}
the bimodule  $\Omega_{u} ^1 (\Ua)$ extends, by means of the graded Leibniz rule,
to the universal graded differential algebra with $d^2=0$. This leads to
higher order differential calculi. Another universal
extension with $d^N=0$, still for the case of binary (unital) algebras,
has been considered in \cite{DV-K2,DViolette}.
These universal extensions have been provided by means of the
$q-$Lebniz rule, for $q$ being primitive $N-$ degree root of the unity,
i.e. $q=e^{\frac{2\pi i}{N}}$ (see also \cite{Kapr} in this context).
However, the so-called $N-$ary case ($d^N=0$) seems also to be specially well
adopted for $N-$ary algebras. Various constructions of higher order differentials
for ternary algebras will be a subject of our future investigation.
\vskip 0.4cm
\indent
{\bf Acknowledgement}
The authors wish to express their thanks to Michel Dubois-Violette for his
critical reading of the manuscript and many enlightening remarks.
%\newpage

\end{document}